\documentclass[submission,copyright]{eptcs}
\usepackage{subfiles}

\providecommand{\event}{HSVC 2014} % Name of the event you are submitting to
\usepackage{breakurl}              % Not needed if you use pdflatex only.
\usepackage{listings} % Source Code listings.
\usepackage{graphicx} % Allows inclusion of JPEGs.
\usepackage{verbatim} % Allows begin{comment}
\usepackage{ifthen}
\usepackage{xspace}
\usepackage{amssymb, amsmath, stmaryrd, url}
\usepackage{tikz}
\usetikzlibrary{arrows}
\usepackage{pstricks,pst-node,pst-tree}
\usepackage{pifont}
\usepackage{varwidth}
\usepackage{hhline}
\usepackage{multirow}
\usepackage{rotating}
\usepackage{latexsym}
\usepackage{etex}
\usepackage{todonotes}
\usepackage{wrapfig}
\usepackage{upgreek}

\newcommand{\ignore}[1]{}

\newcommand{\abbrevfullstop}[1]{%
  \ifthenelse{\equal{#1}{.}}{.}{%
    \ifthenelse{\equal{#1}{,} \OR \equal{#1}{;} \OR \equal{#1}{'} %
            \OR \equal{#1}{?} \OR \equal{#1}{!} %
            \OR \equal{#1}{)} \OR \equal{#1}{]} %
            \OR \equal{#1}{~}}{.#1}%
    {.\ #1}}}

\newcommand{\widen}{\mathbin{\nabla}}
\newcommand{\widenupto}[1]{\mathbin{\nabla_{#1}}}
\newcommand{\lub}{\mathbin{\sqcup}}
\newcommand{\meet}{\mathbin{\sqcap}}

\newcommand{\qquote}[1]{\lquote #1 \rquote}

\def \lquote    {[\![}
\def \rquote    {]\!]}

\newcommand{\pcode}[2][\codesize]{
    \fbox{
    \begin{minipage}{0.45\linewidth}
    #1
    \begin{tabbing}
    xx \= xx \= xx \= xx \= xx \= xx \= xx \= xx \= \kill
    #2
    \end{tabbing}
    \end{minipage}
    }
  }

\newpsobject{showgrid}{psgrid}{subgriddiv=1,griddots=10,gridlabels=6pt}

\newcommand{\ppcode}[2][\small]{
\begin{minipage}{0.45\linewidth}
 #1
\begin{tabbing}
xx \= xx \= xx \= xx \= xx \= xx \= xx \= xx \= \kill
#2
\end{tabbing}
\end{minipage}
}

\newcommand{\statepair}[2]{\mbox{$\langle \mbox{#1} \mid \mbox{#2} \rangle$}}

\newcommand{\clppred}[2]{\mbox{\textsf{#1($#2$)}}}

\newcommand{\crossmark}{\textcolor{red}{\ding{55}}}  % Crossmark

\newcommand{\showcomment}[2]{\ifthenelse{\equal{\enablecomments}{true}}{#1}{#2}}
\newcommand{\enablecomments}{false}
\title{Verification of Programs by Combining Iterated Specialization with Interpolation}
\author{
  Emanuele De Angelis and   Fabio Fioravanti
  \institute{DEC, University G. d'Annunzio,  Pescara, Italy}
  \email{\{emanuele.deangelis, fioravanti\}@unich.it}
  \and Jorge A. Navas
  \institute{NASA Ames Research Center, Moffett Field, USA}
  \email{jorge.a.navaslaserna@nasa.gov}
  \and Maurizio Proietti
  \institute{IASI-CNR, Rome, Italy}
  \email{maurizio.proietti@iasi.cnr.it}
}

\def\titlerunning{Verification of Programs by Combining Iterated Specialization with Interpolation}
\def\authorrunning{E. De Angelis, F. Fioravanti, J. A. Navas \& M. Proietti}

\begin{document}
\maketitle

\begin{abstract}
We present a verification technique for program safety that combines
{\em Iterated Specialization} and {\em Interpolating Horn Clause
  Solving}.  Our new method composes together these two
techniques in a modular way by exploiting the common Horn Clause
representation of the verification problem.  The Iterated
Specialization verifier transforms an initial set of verification
conditions by using unfold/fold equivalence preserving transformation
rules.  During transformation, program invariants are discovered by
applying widening operators.  Then the output set of specialized
verification conditions is analyzed by an Interpolating Horn Clause
solver, hence adding the effect of interpolation to the effect of
widening.  The specialization and interpolation phases can be
iterated, and also combined with other transformations that change the
direction of propagation of the constraints (forward from the program
preconditions or backward from the error conditions).  We have
implemented our verification technique by integrating the VeriMAP
verifier with the FTCLP Horn Clause solver, based on Iterated
Specialization and Interpolation, respectively.  Our experimental
results show that the integrated verifier improves the precision of
each of the individual components run separately.
% and is competitive with
%state-of-the-art Horn Clause-based tools.

\end{abstract}

%\subfile{1intro.tex}
%\subfile{2example.tex}
%\subfile{3method.tex}
%\subfile{4experiments.tex}
%\subfile{5related.tex}
%\subfile{6conclusions.tex}

%\documentclass[14-HCVS-FLoC.tex]{subfiles}
%\begin{document}

\section{Introduction}
\label{sec:intro}

%\textbf{Jorge:} I would like to avoid mentioning FTCLP as much as
%possible to keep the paper general. We could use the term
%``interpolating horn-clase based verifier (IHCV)'' or something like
%that. Then, only for experiments we mention FTCLP.
%
%\emanuele{TBD}
%\fabio{TBD}
%\maurizio{TBD}
%\jorge{TBD}
%
%Motivations: Many methods for HC-based
%verification\cite{disjunctive-intps,GrebenshchikovLPR12,JaffarMNS12}
%(e.g., based on invariant generation/interpolation) each with its own
%merits, Program Transformation\cite{FPP14} can be used for combining
%them and exploiting synergy.

{\em Constraint Logic Programming}\footnote{In this paper we use interchangeably
the terminology ``Constraint Logic Program" and ``Constrained Horn Clauses" \cite{BjornerMR12}.}
(CLP)~\cite{JaffarL87} is
becoming increasingly popular
as a logical basis for developing methods and tools for
software verification~(see, for instance, \cite{BjornerMR12,De&14c,GrebenshchikovLPR12,JaffarMNS12,disjunctive-intps,PeraltaG98}).
Indeed, CLP provides a suitable formalism for expressing {\em verification conditions} that
guarantee the correctness of imperative, functional, or concurrent programs and,
moreover, constraints are very useful for encoding properties of data
domains such as integers, reals, arrays, and heaps  \cite{Bag&08,Du&13,mathsat5,MouraB08}.
An advantage of using a CLP representation for verification problems
is that we can then combine reasoning techniques and constraint solvers
based on the common logical language \cite{GaK14}.
In this paper we will show that, by exploiting the CLP representation,
we can combine in a modular way 
a verification technique based on CLP {\em specialization} \cite{De&14c}
and a verification technique
based on interpolation \cite{GangeNSSS13}.

The use of CLP program specialization, possibly integrated with abstract interpretation,
has been proposed in various verification techniques 
\cite{AlbertGHP07,De&14c,PeraltaG98,PueblaAH06}.
In particular,
{\em iterated specialization} \cite{De&14c}  makes use of {\em unfold/fold}
CLP program transformations to specialize a given set 
of verification conditions with respect to the constraints representing
the {\em initial} and {\em error} configurations.
The objective of the specialization transformation is to derive an equivalent
set of verification conditions represented as a finite set of constrained 
facts from which one can immediately infer program correctness or incorrectness.
Since the verification problem is undecidable in general,
the specialization strategy employs a suitable {\em generalization strategy}
to guarantee the termination of the analysis.
Generalization consists in replacing a clause \mbox{{\tt  H :- c,B}}
by the new clause {\tt H :- c,G},
where {\tt G} is defined by the clause {\tt G :- d,B} and {\tt d} is a constraint
entailed by {\tt c}, and then specializing {\tt G}, instead of {\tt H}.
Most generalization strategies are based on {\em widening}
operators like the ones introduced in the field of {abstract interpretation} \cite{Cousot_POPL77}.
Unfortunately, the use of generalization can lead to a loss of precision,
that is, to a specialized set of verification conditions which 
we are neither able to prove satisfiable nor to prove unsatisfiable
because of the presence of recursive clauses.
Precision can be improved by iterating the specialization process and
alternating the propagation of the constraints of the initial configuration (forward propagation)
and of the error configuration (backward propagation).

{\em Interpolation}~\cite{interpolant} is a technique that has been
proposed to recover precision losses from abstractions, including those
from widening (see, for instance, \cite{AGCSAS12,DAGGER08}).  Having
found a spurious counterexample (that is, a false error detection),
one can compute an interpolant formula that represents a set of states
from which the spurious counterexample cannot be generated, and by
combining widening with interpolation one can recover precision of the
analysis by computing an overapproximation of the reachable states
that rules out the counterexample.  Interpolation (without any
combination with widening) has also been used by several CLP
verification techniques (see, for instance,
\cite{GangeNSSS13,GrebenshchikovLPR12,JaffarSV09,Duality,disjunctive-intps}).
However, to the best of our knowledge, no CLP-based verifier combines
widening and interpolation in a nontrivial way.

Among the various CLP-based techniques, {\em Interpolating Horn
  Clause} (IHC) solvers~\cite{GangeNSSS13} enhance the classical
top-down (i.e., goal oriented) execution strategy for CLP with the use
of interpolation.  During top-down execution suitable interpolants are
computed, and when the constraints accumulated during the execution of
a goal imply an interpolant associated with that goal, then it is
guaranteed that the goal will fail and the execution can be stopped.
Thus, the use of interpolants can achieve termination in the presence
of potentially infinite goal executions.  Therefore, IHC solvers can
be used to verify safety properties by showing that error conditions
cannot be reached, even in the presence of infinite symbolic program
executions.

In this paper we propose a CLP-based verification technique that
combines widening and interpolation by composing in a modular way
specialization-based verification with an IHC solver.  Our technique
first specializes the CLP clauses representing the verification
conditions with respect to the constraints representing the initial
configuration, hence deriving an equivalent set of CLP clauses that
incorporate constraints derived by unfolding and generalization
(including, in particular, loop invariants computed by widening).
Next, the IHC solver is applied to the specialized CLP clauses, hence
adding the effect of interpolation to the effect of widening.  In the
case where the interpolation-based analysis is not sufficient to check
program correctness, the verification process proceeds by specializing
the CLP clauses with respect to 
the constraints representing 
the error configuration, and then
applying the IHC solver to the output program.  Since
program specialization preserves equivalence with respect to the
property of interest, we can iterate the process consisting in
alternating specialization (with respect to the initial or error
configuration) and interpolation in the hope of eventually deriving
CLP clauses for which we are able to prove correctness or
incorrectness.

We have implemented our verification technique by integrating 
FTCLP, an IHC solver~\cite{GangeNSSS13}, within VeriMAP,
a verifier based on iterated specialization \cite{DeAngelisFPP14}.  
The experimental results show
that the integrated verifier improves the precision of each of its
individual components.
% and is competitive with state-of-the-art
%CLP-based verifiers.

%\vspace{-7mm}

%\newpage

%\end{document}

%\documentclass[14-HCVS-FLoC.tex]{subfiles}
%\begin{document}
\section{An Introductory Example}
\label{sec:example}

Consider the C program in Figure~\ref{example}(a) where the symbol $*$
represents a non-deterministic choice. A possible inductive
invariant that proves the safety of
this program is $x \geq 1 \wedge y \geq 0 \wedge x \geq
y$.%%  since the following three conditions hold:

%% \begin{itemize}

%% \item $x=1 \wedge y=0 \models x \geq 1 \wedge y \geq 0 \wedge x \geq
%%   y$ \hfill (\textsf{initiation})

%% \item $x \geq 1 \wedge y \geq 0 \wedge x \geq y \wedge x'=x+y \wedge
%%   y'=y+1 \models x' \geq 1 \wedge y' \geq 0 \wedge x' \geq y'$
%%   \hfill (\textsf{consecution})

%% \item $x \geq 1 \wedge y \geq 0 \wedge x \geq y \models x \geq y$
%%   \hfill (\textsf{safety})
%% \end{itemize}

Figure~\ref{example}(b) shows the set of CLP clauses that represent
the verification conditions for our verification problem. We will
explain in Section~\ref{sec:method} how to generate automatically
those conditions, but very roughly each basic block in the C program is
translated to a CLP clause and each assertion is negated. Note that
the loop is translated into the recursive predicate \textsf{new3}. 
The key property of this translation is that the program in
Figure~\ref{example}(a) is safe iff the CLP
predicate \textsf{unsafe} in Figure~\ref{example}(b) is unsatisfiable, 
that is, all derivations of \textsf{unsafe}
lead to failure.

\begin{figure*}[h]
\begin{tabular}[t]{|c|c|c|}

\hline
%%%%%%%%%%%%%%%%%%%%%%%%%%%%%%%%%%%%%%%%%%%%
\ppcode{
  \textbf{int} $x$ = $1$; \\
  \textbf{int} $y$ = $0$; \\
  \textbf{while} ($*$) \{ \\
  \>   $x$ = $x+y$; \\
  \>   $y$ = $y+1$; \\
  \}\\
  \textbf{assert}( $x>=y$);
} 
&
%%%%%%%%%%%%%%%%%%%%%%%%%%%%%%%%%%%%%%%%%%%%
\ppcode{
  \textsf{unsafe}        :- \textsf{new2}. \\
  \textsf{new2}          :- $X=1$, $Y=0$, \textsf{new3}($X,Y,\_$). \\
  \textsf{new3}($X,Y,\_$)   :-  \textsf{new4}($X,Y,\_$). \\
  \textsf{new4}($X,Y,C$)   :- $X1=X+Y$,  \\
  \>  $Y1=Y+1$, $C \geq 1$,  \textsf{new3}($X1,Y1,C$). \\
  \textsf{new4}($X,Y,C$)   :-  $C \leq 0$, \textsf{new6}($X,Y,C$). \\
  \textsf{new6}($X,Y,C$)   :-  $D=1$, $(X-Y) \geq 0$ , \\
  \> \textsf{new7}($D,X,Y,C$). \\
  \textsf{new6}($X,Y,C$)   :-  $D=0$, $(X-Y) \leq -1$, \\
  \> \textsf{new7}($D,X,Y,C$). \\
  \textsf{new7}($D,\_,\_,\_$) :- $D=0$. 
} 
&
\ppcode{
\textsf{unsafe}      :- \textsf{new2}. \\
\textsf{new2}        :- $X=1$, $Y=0$,  \textsf{new4}($X,Y$). \\
\textsf{new4}($X,Y$) :- $X=1$, $Y=0$, \\
\>   $Y1=1$, $X1=1$, \textsf{new5}($X1,Y1$).\\

\textsf{new5}($X,Y$) :-  $X=1$, $Y \geq 0$,  \textsf{new8}($X,Y$).\\

\textsf{new8}($X,Y$) :- $X=1$, $X1=Y+1$,  \\
 \> $X1 \geq 1$, $Y1=X1$, \textsf{new9}($X1,Y1$).\\
 
\textsf{new8}($X,Y$) :- $X=1$, $Y \geq 0$, \textsf{new10}($X,Y$).\\

\textsf{new10}($X,Y$):- $X=1$, $Y \geq 2$.\\

\textsf{new9}($X,Y$) :-  $X \geq 1$,$Y \geq 0$, \textsf{new13}($X,Y$).\\

\textsf{new13}($X,Y$):- $X1=X+Y$, $Y1=Y+1$, \\ 
\>  \textsf{new9}($X1,Y1$).\\
\textsf{new13}($X,Y$):- $X \geq 1$,$Y \geq 0$, \textsf{new15}($X,Y$).\\ 

\textsf{new15}($X,Y$):- $X \geq 1$, $(X-Y) \leq -1$. \\

} \\
%%%%%%%%%%%%%%%%%%%%%%%%%%%%%%%%%%%%%%%%%%%%
\mbox{(a)} & \mbox{(b)} & \mbox{(c)} \\
%%%%%%%%%%%%%%%%%%%%%%%%%%%%%%%%%%%%%%%%%%%%
\hline 

\end{tabular}

\caption{(a) C program, (b) CLP clauses representing the verification
  conditions, and (c) CLP clauses after specialization. \label{example}}
\end{figure*}

Note that neither a top-down nor a bottom-up evaluation of the CLP
program in Figure~\ref{example}(b) will terminate, essentially due to
the existence of the recursive predicate \textsf{new3}. To avoid this problem, our
method performs a transformation based on \emph{program
specialization} \cite{De&14c}. We postpone the details of this
transformation to Section~\ref{sec:method}, but the key property of this
transformation is that it is \emph{satisfiability-preserving}. That is,
\textsf{unsafe} is satisfiable in the input program iff it is satisfiable 
in the transformed one. The benefits of this transformation
come from the addition of new constraints (e.g., using \emph{widening}
techniques) that might make a top-down or bottom-up evaluation
terminate. We show the result of this transformation in
Figure~\ref{example}(c). For clarity, we have removed from the
transformation irrelevant arguments and performed some constraint
simplifications.

The clause of predicate \textsf{new2} encodes the initial conditions. The
transformation introduces new definitions \textsf{new4} and
\textsf{new8} after unrolling twice \textsf{new3}.  The predicate
\textsf{new9} encodes the loop just after these two unrolls. Note that
our transformation inserts two additional constraints $X \geq 1$ and
$Y \geq 0$ that will play an essential role later.
Although this transformed set of clauses might appear simpler to solve
the predicate \textsf{unsafe}, it is still defined by recursive clauses with
constrained facts. Even CLP systems with
tabling~\cite{GuzmanCHS12,Codognet95} will not terminate here.

Instead, consider an IHC solver following the approach of
\emph{Failure Tabled CLP} (FTCLP)~\cite{GangeNSSS13}.
% . Since these verifiers
%(\cite{Duality,disjunctive-intps,GrebenshchikovLPR12,JaffarMNS12})
%can significantly vary from each other we will use in this paper
In a nutshell, FTCLP augments tabled CLP~\cite{GuzmanCHS12,Codognet95} by computing
\emph{interpolants} whenever a failed derivation is encountered
during the top-down evaluation of the CLP program. If a call of predicate $p$
has been fully executed 
and has not produced any solution then its
interpolants denote the conditions under which the execution of $p$
will always lead to failure.
During the execution of a recursive predicate $p$ the top-down evaluation
might produce multiple copies of $p$, each one with different
constraints $c$ originated from the unwinding of the recursive clauses
for $p$. Given two
copies $p'$ and $p''$ with constraints $c'$ and $c''$, respectively,
and being $p''$ a descendant of $p'$, the tabling mechanism will check
whether $p''$ is \emph{subsumed} by $p'$ (i.e., $ c'' \sqsubseteq
c'$). If this is the case, the execution can be safely stopped at
$p''$.
In tabling, rather than using $c'$ it is common to use weaker
constraints (in the logical sense) in order to increase the likelihood
of subsumption. These weaker constraints are often called \emph{reuse
  conditions}.
The major difference with respect to standard tabled CLP is that FTCLP
uses the interpolants as reuse conditions while tabled CLP uses
\emph{constraint projection}. A key insight is that when FTCLP is used
in the context of verification its tabling mechanism using
interpolants as reuse conditions resembles a verification
algorithm using interpolants to produce \emph{candidates} and keeping
those which can be proven to be inductive invariants.

Coming back to our CLP program in Figure~\ref{example}(b). If we 
run FTCLP on that set of clauses, the top-down execution of
\textsf{unsafe} will not terminate. To understand why, let us focus on
the predicate \textsf{new3}, initially with constraints $X=1,
Y=0$. From \textsf{new3} we reach \textsf{new4} which has two
clauses. From the execution of the first clause, we will reach again
\textsf{new3} ($\Pi_{1}$ in Table~\ref{example-paths}), but this time
with constraints $X=1, Y=0, X1=X+Y,Y1=Y+1$ (we ignore constraints over
$C$ since they are irrelevant). To avoid an infinite loop, the tabling
mechanism will freeze here its execution and backtrack to the nearest
choice point which is the second clause for \textsf{new4}. From here,
we will reach \textsf{new6} which has also two clauses. The first
clause for \textsf{new6}, executed by $\Pi_{2}$, has the constraint $X
-Y \geq 0, D=1$ which is consistent with $X=1, Y=0,
X1=X+Y,Y1=Y+1$. The execution follows \textsf{new7} which will fail
because the constraint $D=0$ cannot be satisfied. The second clause of
\textsf{new6} executed in $\Pi_{3}$ has no solution either because $D=0,
X - Y \leq -1$ is also false. Very importantly, FTCLP will generate
interpolants from these two failed derivations. The key interpolant is
generated from the second clause of \textsf{new6} ($\Pi_{3}$) which when
propagated backwards to \textsf{new3} is $X \geq Y$. Note that this is
an essential piece of information, but unfortunately it is not enough
for tabling to stop permanently since this constraint alone is not
sufficient to subsume the descendants of \textsf{new3}. That is, $X
\geq Y \wedge X1=X+Y \wedge Y1=Y+1 \not\models X1 \geq Y1$. Therefore,
FTCLP will unfreeze the execution of \textsf{new3} and repeat the
process. Unfortunately, FTCLP will not terminate.
%because it cannot add to the subsumption test constraints such as $X
%\geq 1$ and $Y \geq 0$.

\begin{table}[t]
\begin{small}
\begin{tabular}{|c|c|}
\hline
\textsf{CLP constraints for Figure~\ref{example}(b)} & \textsf{CLP constraints for Figure~\ref{example}(c)} \\
\hline
\begin{minipage}{0.53\linewidth}
\begin{tabular}{@{\hspace{-1mm}}p{0.5mm}ll}
 $\Pi_{1}$: & $~~\ldots$, $X=1$,$Y=0$, \underline{\textsf{new3}($X,Y,\_$)}, \textsf{new4}($X,Y,\_$), \\ 
          & $X1=X+Y$, $Y1=Y+1$, \underline{\textsf{new3}($X1,Y1,\_$)}   \\
%%%%%%%%%%%%%
 $\Pi_{2}$: & $~~\ldots$, $X=1$,$Y=0$, \textsf{new3}($X,Y,\_$), \textsf{new4}($X,Y,\_$),  \\
         & \textsf{new6}($X,Y,\_$),  $X-Y \geq 0$, $D=1$, \textsf{new7}($D,X,Y,\_$), $D=0$  \\ 
%%%%%%%%%%%%%
 $\Pi_{3}$: & $~~\ldots$, $X=1$,$Y=0$, \textsf{new3}($X,Y,\_$), \textsf{new4}($X,Y,\_$),   \\
        & \textsf{new6}($X,Y,\_$), $X-Y \leq -1$, $D=0$, \textsf{new7}($D,X,Y,\_$), $D=0$   \\ 
\end{tabular}
\end{minipage} & 
\begin{minipage}{0.4\linewidth}
\begin{tabular}{@{\hspace{-0.5mm}}p{0.5mm}ll}
 $\Pi_{1'}$: & $~~~X=2$,$Y=2$, \underline{\textsf{new9}($X,Y$)}, $X\geq1, Y \geq 0$, \\
          & \textsf{new13}($X,Y$),$X1=X+Y$, $Y1=Y+1$, \\
          & \underline{\textsf{new9}($X1,Y1$)}   \\ 
%%%%%%%%%%%%%
 $\Pi_{2'}$: & $~~~X=2$,$Y=2$, \textsf{new9}($X,Y$), $X\geq1, Y \geq 0$, \\
          & \textsf{new13}($X,Y$), $X\geq1, Y \geq 0$, \textsf{new15}($X,Y$)   \\ 
          & $X \geq 1, X - Y \leq -1$ \\
%%%%%%%%%%%%%
\end{tabular}
\end{minipage} \\
\hline
\end{tabular}
\end{small} 
\caption{Executions of the CLP clauses from Figures~\ref{example}(b)-(c)\label{example-paths}}
\end{table}

Let us now consider the set of clauses in Figure~\ref{example}(c). Let
us repeat the same process running FTCLP on the transformed
program. Let us focus on \textsf{new9}, which encodes the loop after
two unrolls. For clarity of presentation the execution of
\textsf{new9}, shown in $\Pi_{1'}$, starts with the constraints
$X=2,Y=2$ which are produced by projecting the constraints accumulated
from \textsf{new2} to \textsf{new9} onto $X$ and $Y$. Since there is a
cycle when \textsf{new9}($X1,Y1$) is reached the tabling mechanism
will freeze its execution.
% but this time with constraints $\ldots, X \geq 1, Y \geq 0, X1=X+Y,
% Y1=Y+1$.
Note that the transformation inserted the constraints $ X \geq 1$ and
$Y \geq 0$. This was achieved by applying generalization via widening
during the unfolding of the clause.
From the second clause of \textsf{new13}, shown in $\Pi_{2'}$, the
top-down evaluation will eventually fail. Again, it will generate an
interpolant that after backwards propagation to \textsf{new9} is $X
\geq Y$. Now, it can be proven that the descendant of \textsf{new9}
can be safely subsumed because $ X \geq Y , X \geq 1, Y \geq 0,
X1=X+Y, Y1=Y+1 \models X1 \geq Y1$, and therefore, FTCLP will
terminate proving that \textsf{unsafe} is unsatisfiable.
The magic here is originated from the fact that the transformation
produced the invariants $X \geq 0$ and $Y \geq 1$ (\emph{widening})
while FTCLP produced the remaining part $X \geq Y$
(\emph{interpolation}) which together form the desired safe inductive
invariant.

Finally, it is worth mentioning that we do not claim that this program
cannot be proven safe by other methods using more sophisticated
interpolation algorithms (e.g.~\cite{AlbarghouthiLGC12}) and/or
combining with other techniques such as predicate abstraction
(e.g.,~\cite{GrebenshchikovLPR12}). Instead, we would like to stress
how we can have the same effect than a specialized algorithm using
widening and interpolation in a much less intrusive and completely
modular manner by exploiting the fact that both methods share the same
CLP representation.

%\end{document}

% \documentclass[14-HCVS-FLoC.tex]{subfiles}

% to be moved to main latex file
%\newcommand{\down}{\rule{0mm}{3.5mm}}
%\newcommand{\band}{\rule{0mm}{3mm}}

%\newcommand{\usc}{\hspace{-.2mm}\_\hspace{.35mm}} 

% \begin{document}

\section{Verification based on Iterated Specialization and Interpolation}
\label{sec:method}

%% In order to show how our verification method based on iterated
%% specialization and interpolation works, let us consider again the
%% following imperative program $P$, already shown in
%% Fig.~\ref{example}~(a):

%% \smallskip

%% \hspace{0mm}$\texttt{int}~~x=1;$
%% \makebox[1mm][l]{}
%% $\texttt{int}~~y=0;$
%% \makebox[1mm][l]{}

%% $\ell_{0}$: \texttt{while}~~$(*)~~\{\,x=x+y; ~~y=y+1;\,\}$
%% \makebox[1mm][l]{}

%% \smallskip
%% \noindent

In order to show how our verification method based on iterated
specialization and interpolation works, let us consider again the
program shown in Fig.~\ref{example}~(a).
We want to prove safety of program $P$ with respect to the initial
configurations satisfying $\varphi_{\mathit{init}}(x,y)
=_{\mathit{def}} x\!=\!1 \ \wedge$ $y\!=\!0$, and the error
configurations satisfying $\varphi_{\mathit{error}}(x,y)
=_{\mathit{def}} x\! <\! y$. That is, we want to show that, starting
from any values of $x$ and $y$ that satisfy
$\varphi_{\mathit{init}}(x,y)$, after every terminating execution of program~$P$,
the new values of $x$ and $y$ do not satisfy $\varphi_{\mathit{error}}(x,y)$.

\smallskip
Our verification method consists of the following steps.
\smallskip
\\ (\textit{Step 1 - CLP Encoding}) First, we encode the safety
verification problem using a CLP program $I$, called the
\emph{interpreter}, which defines a predicate $\texttt{unsafe}$. We
have that program $P$ is safe iff $I \not\models \texttt{unsafe}$. 
\showcomment{++++ intro to CLP is missing}{}
\smallskip
\\ (\textit{Step 2 - Verification Conditions Generation}) Then, by
specializing $I$ w.r.t.~(a CLP encoding of) program $P$, we generate a
set \textit{VC} of CLP clauses representing the \emph{verification
  conditions} for $P$. Specialization preserves safety, i.e., 
  $I \models \texttt{unsafe}$ iff $\textit{VC} \models \texttt{unsafe}$.
  \showcomment{ +++The specialization of CLP programs is
  performed by applying equivalence-preserving unfold/fold
  transformation rules, guided by strategies.}{}
\smallskip
\\ (\textit{Step 3 - Constraint Propagation}) Next, we apply CLP
specialization again and we propagate the constraints occurring in the
initial and error conditions, thereby deriving a new set $\textit{VC}'$
of verification conditions. Also this specialization step preserves safety, i.e.,
$\textit{VC} \models \texttt{unsafe}$ iff $\textit{VC}' \models \texttt{unsafe}$. 
During this step, in order to guarantee
termination, we generate inductive loop invariants by using
generalization operators based on widening and
convex hull~\cite{Fi&13a}.
\smallskip
\\ (\textit{Step 4 - Interpolating Verification}) We consider the
verification conditions obtained after constraint propagation.  If
their satisfiability cannot be decided by simple syntactic checks
(e.g., emptiness of the set of CLP clauses or absence of constrained facts), we
apply the interpolating Horn Clause solver. 

In the case where the IHC
solver is not able to provide a definite answer, we \emph{reverse}
the program encoding the VC's by exchanging the direction of the
transition relation and the role of the initial and error conditions,
then we \emph{iterate} the constraint propagation and satisfiability
check (go to Step 3).

%The sequence of steps performed by our method is \\
%{\small
%\textit{CLP Encoding; Verification Condition Generation; (Constraint Propagation; Interpolating Verification)*}
%}

%+++++Unfortunately, it is not possible to check by direct evaluation
%whether or not the atom \texttt{unsafe} is a consequence of the above CLP clauses.
%Indeed, the  evaluation of the query \texttt{unsafe} using the standard
%top-down strategy gets into an infinite loop.
%Tabled evaluation~\cite{CuW00} does not terminate either, as infinitely many tabled atoms are generated. Analogously, bottom-up evaluation is 
%unable to return an answer,
%because it has to generate infinitely many facts for \texttt{new1} and \texttt{new2}
% for deriving that \texttt{unsafe} is not a consequence of the
%given CLP clauses.

\medskip
\noindent

Of course, due to undecidability of safety, our verification method
might not terminate.  However, as experimentally shown in
Section~\ref{sec:exp}, the combination of program specialization and
interpolating verification is successful in many examples and it is
synergistic, in the sense that it improves over the use of program
specialization and interpolating verification alone.
We now describe in more detail the techniques applied in each step
listed above.

%\vspace{-3mm}
\subsection{Encoding safety problems of imperative programs using CLP}
\label{subsec:method-step1}
%\vspace{-1mm}

As already mentioned, the safety verification problem can be encoded as a CLP program $I$,
called the interpreter. 
%such that safety of the program is defined in terms of 
%the least model semantics of $I$.
%
We now show, through an example, how to derive a set of CLP clauses that encode 
(i) the semantics of the programming language 
in which the program under verification is written, 
(i) the program under verification itself, and 
(iii) the proof rules for the considered safety property.
The extension to the general case is straightforward.

\smallskip
\noindent
\textbf{Encoding the semantics of the programming language.} 
The  semantics of the imperative language 
can be encoded 
as a transition relation
from any configuration of the imperative program
to the next configuration,
by using the predicate~$\texttt{tr}$.
Below we list the clauses of~$\texttt{tr}$
for 
\noindent
(i)~assignments (clause~1), 
\noindent
(ii)~conditionals (clauses~2 and~3), and 
\noindent
(iii)~jumps (clause 4).

{\small
\vspace{.5mm}
\smallskip
\noindent    % assignment
1.~\texttt{tr(cf(cmd(L,asgn(X,expr(E))),Env), cf(cmd(L1,C),Env1))} {\tt{:-}}
\nopagebreak

\vspace{.2mm}\hspace{25mm}
\texttt{eval(E,Env,V), }\texttt{update(Env,X,V,Env1), nextlab(L,L1), at(L1,C). }

\vspace{.6mm}\noindent    % if-then-else true
2.~\texttt{tr(cf(cmd(L,ite(E,L1,L2)),Env), cf(cmd(L1,C),Env))} {\tt{:-}}
\texttt{beval(E,Env),}%\nopagebreak
%
%\vspace{.2mm}\hspace{35mm}
\texttt{at(L1,C).}

\vspace{.6mm}\noindent   % if-then-else false
3.~\texttt{tr(cf(cmd(L,ite(E,L1,L2)),Env), cf(cmd(L2,C),Env))} {\tt{:-}}
\texttt{beval(not(E),Env),}%\nopagebreak
%
%\vspace{.2mm}\hspace{35mm}
\texttt{at(L2,C).}

\vspace{.6mm}\noindent       % goto
4.~\texttt{tr(cf(cmd(L,goto(L)),Env), cf(cmd(L,C),Env))} {\tt{:-}} 
\texttt{at(L,C).} 

}

\smallskip

\noindent
%The term
%$\texttt{asgn(X,expr(E))}$ encodes the assignment of the value of the expression {\tt{E}} to the variable {\tt{X}}.
%$x\!=\!e$ (in particular, 
%$X$ encodes the identifier $x$ and $E$ encodes the expression $e$).
%the value of the 
%expression $\texttt{E}$ to the variable~$X$ of the form
%\mbox{$x\!=\!e$}. 
The term $\texttt{cf(cmd(L,C),Env)}$ encodes the configuration consisting of the
command~$\texttt{C}$ with label $\texttt{L}$
and the environment  $\texttt{Env}$.
%The term $\texttt{asgn(X,expr(E))}$ encodes the assignment of the value of the expression {\tt{E}} to the variable {\tt{X}}.
%The term $\texttt{ite(E,L1,L2)}$ encodes the if-then-else command
%jumping to either label {\tt{L1}} or label {\tt{L2}}, 
%according the Boolean value of the expression {\tt{E}}.
%The term $\texttt{goto(L)}$ encodes the jump to label {\tt{L}}.
%
%
The predicate $\texttt{eval(E,Env,V)}$ computes the value~$\texttt{V}$ of the 
expression~$\texttt{E}$ in the environment~$\texttt{Env}$.
The predicate $\texttt{beval(E,Env)}$ holds if the Boolean expression
$\texttt{E}$ is true in the  environment~$\texttt{Env}$.
The predicate $\texttt{at(L,C)}$ binds to $\texttt{C}$ the command 
with label~$\texttt{L}$.
The predicate $\texttt{nextlab(L,L1)}$ binds to~$\texttt{L1}$ the label of
the command 
that is written immediately after the command 
with label~$\texttt{L}$.
The predicate $\texttt{update(Env,X,V,Env1)}$ updates 
the  environment~$\texttt{Env}$ by binding the variable~$\texttt{X}$ 
to the value~$\texttt{V}$, thereby constructing 
a new  environment~$\texttt{Env1}$.

\medskip
%\smallskip
\noindent
\textbf{Encoding the imperative program.}  The imperative program $P$
is encoded by a set of constrained facts for the $\texttt{at}$
predicate, as follows.  First, we translate program $P$ to the
sequence of labelled commands $\ell_{0} \ldots \ell_{h}$. Then, we
introduce the CLP facts $\{5-9\}$ which encode those commands.

\smallskip

{\small

\begin{tabular}{ l l }

\makebox[3mm][l]{}$\ell_{0}$: $\texttt{if}~(*)~~   
\ell_{1}
\texttt{~else~}\ell_{h}$\,;\nopagebreak  
& 
5.~$\texttt{at(0,ite(nondet,1,h))}$.\nopagebreak

\\
\makebox[3mm][l]{}$\ell_{1}$: $x=x+y$\,;\nopagebreak
&
6.~$\texttt{at(1,asgn(int(x),expr(plus(int(x),int(y)))))}$.\nopagebreak

\\
\makebox[3mm][l]{}$\ell_{2}$: $y=y+1$\,;\nopagebreak
&
7.~$\texttt{at(2,asgn(int(y),expr(plus(int(y),int(1)))))}$.\nopagebreak

\\
\makebox[3mm][l]{}$\ell_{3}$: $\texttt{goto}~ \ell_{0}$\,;
&
8.~$\texttt{at(3,goto(0))}$.\nopagebreak

\\
\makebox[3mm][l]{}$\ell_{h}$: $\texttt{halt}\,;$
&
9.~$\texttt{at(h,halt)}$. 

\end{tabular}
}

%%%%
\medskip
%\smallskip
\noindent
\textbf{Encoding safety as reachability of configurations.}
Finally, we write the CLP clauses defining the predicate 
$\texttt{unsafe}$ that holds if and only if there exists an
 execution of the program~$P$ that leads 
from an initial configuration to an error configuration.

{\small

\smallskip
\noindent
10.~$\texttt{unsafe}$ {\tt{:-}} $\texttt{initConf(X), reach(X).}$

\noindent
11.~$\texttt{reach(X)}$ {\tt{:-}} $\texttt{tr(X,X1), reach(X1).}$

\noindent
12.~$\texttt{reach(X)}$ {\tt{:-}} $\texttt{errorConf(X).}$

\noindent
13.~\texttt{initConf(cf(cmd(0,C),} 
\texttt{[[int(x),X],[int(y),Y]]))} 
%\hspace{15mm}
{\tt{:-}} \texttt{X=1,\,Y=0,\,at(0,C)}.

\noindent
14.~\texttt{errorConf(cf(cmd(h,halt),} 
\texttt{[[int(x),X],[int(y),Y]]))} {\tt{:-}} \texttt{X<Y}.

}

\smallskip
\noindent
Note that in clauses 13 and 14, 
the second component of the
configuration term $\texttt{cf}$
encodes the environment 
as a list 
$\texttt{[[int(x),X],[int(y),Y]]}$
that provides the bindings for the 
program variables~$x$ and $y$, respectively.
%
%
%\smallskip
%\noindent
The set of CLP clauses $\{1-14\}$ constitutes the interpreter $I$.
By the correctness of the CLP Encoding~\cite{De&14c} 
we have that program $P$ is safe if and only if $I \not\models\texttt{unsafe}$.

When the CLP program $I$ is run by using the top-down, goal 
directed strategy usually employed by CLP systems, an
execution of the goal $\texttt{unsafe}$ corresponds to a forward
traversal of the transition graph (i.e., a traversal
starting from the initial configuration).
We can also provide an alternative definition of the
$\texttt{reach(X)}$ relation that would generate a backward
traversal of the transition graph (i.e., a traversal
starting from the error configuration).
However, as shown in Section \ref{subsec:method-step5}, the method
alternates between the forward and backward directions, and the
direction used by the first specialization is not very relevant in practice.

More in general, the specialization-based approach is to a large extent
parametric with respect to the definition of the semantics of
the programming language, as long as this is defined by a CLP program.

%\vspace{-3mm}
\subsection{Generation of CLP Verification Conditions}
\label{subsec:method-step2}
%\vspace{-1mm}

The verification conditions associated with the given safety
verification problem, are generated by performing a CLP
specialization based on unfold/fold program transformations \cite{EtG96,Fi&00b}.  During
this step, %which can be seen as an application of the first Futamura
%projection,
we specialize the clauses defining $\texttt{unsafe}$ in the
interpreter $I$ with respect to: (i)~the clauses that define the
predicate~$\texttt{tr}$, and (ii)~the clauses that encode the given
program~$P$.  The output of this program specialization is the set of
verification conditions for $P$: a set of CLP clauses that encode the
unsafety of the imperative program, is equisatisfiable w.r.t. $I$, and
contains no explicit reference to the predicate symbols used for
encoding the transition relation and the program commands. This first
step is performed also in other specialization-based techniques for
program verification (see, for
instance,~\cite{DeAngelisFPP12b,PeraltaG98}.).
In this paper we consider a C-like imperative programming language and proof rules for safety properties only.
However, the specialization-based approach to the generation of verification conditions 
has the advantage of being modular w.r.t. 
(i) the semantics of the programming language 
in which the program under verification is written,
and (ii) the logic used for specifying the property to be verified,
and seems to be reasonably efficient in practice.

The verification conditions for the safety problem we are considering
are shown in Fig.~\ref{example}~(b) and have been briefly described in
Section~\ref{sec:example}.

\showcomment{++++ ADD some comments. e.g. predicates corresponding to loop head, loop exits...  +++ change parameter for VCGEN in order to obtain simpler VC's? ++++}{}

%%%%%%%%%%%%%%%%

\vspace{-3mm}
\subsection{Constraint Propagation by CLP Transformation}
\label{subsec:method-step3}

Constraint propagation is achieved by a CLP specialization algorithm
similar to the one used for the Verification Conditions Generation (Step 2).  The main
difference is that for constraint propagation we use a {\em
  generalization operator} based on widening and convex-hull that in
many cases allows the discovery of useful program invariants~\cite{Fi&13a}.

The specialization algorithm starts off from a set $VC$ of clauses
that include $j\!\geq\! 1$ clauses defining the predicate
\texttt{unsafe}:

\smallskip

$\texttt{unsafe~:-~c$_1\!$(X),\,p$_1$\!(X)}, ~~\ldots,~~ 
\texttt{unsafe~:-~c$_j\!$(X),\,p$_j$\!(X)}$\nopagebreak

\smallskip

\noindent where 
$\texttt{c$_1\!$(X),\ldots,c$_j$\!(X)}$ are constraints and 
$\texttt{p$_1\!$(X),\ldots,p$_j$\!(X)}$ are atoms. 

The specialization algorithm makes use of the {\em unfolding}, 
%{\em clause removal}, 
{\em definition introduction}, and {\em folding}
transformation rules \cite{EtG96,Fi&00b}.

The clauses for \texttt{unsafe} are unfolded, that is, they are
transformed by applying the following unfolding rule: Given a clause
$C$ of the form {\rm \texttt {H\,:-\,c,A}}, let {\rm $\{\texttt{K}_i\,
  \texttt{:-}\, \texttt{c}_i\texttt{,Q}_i \mid i=1, \ldots, m\}$} be
the set of the $($renamed apart$)$ clauses in program $VC$ such that,
for $i=1,\ldots ,m,$ {\rm$\texttt{A}$} is unifiable with {\rm
  ${\texttt K}_i$} via the most general unifier~$\vartheta_i$ and the
constraint {\rm(\texttt{c,c$_i$})}$\,\vartheta_i$ is satisfiable.
Then from clause $C$ we derive the clauses:

\smallskip

$\mathit{(\texttt{H\,:-\,c,c$_i$,Q$_i$})\,\vartheta_i, \ \mbox{ for } i=1,\ldots, m.}$

\smallskip
\noindent
Unfolding propagates the constraints by adding the constraints on atom
$A$ to the constraints of the atom \texttt{Q$_i$} that
occurs in the body of a clause unifying $A$.
By unfolding we may derive a fact for \texttt{unsafe}, and hence the
given program is unsafe.  Alternatively, we may derive an empty set of
clauses for \texttt{unsafe}, and hence we infer safety.  However, in
most cases we will derive a nonempty set of clauses which are not
constrained facts.
%, and in this case the specialization algorithm
%proceeds by introducing new predicate definitions and then folding.
In these cases the specialization algorithm introduces a set \textit{Defs}
of new predicate definitions, 
one for each clause which is not a constrained
fact. 
More specifically, let $E$ be a clause 
derived by unfolding
of the form  
{\rm $\texttt{H(X)\,:-\,e(X,X1),\,Q(X1)}$}, where {\rm\texttt {X}}  and 
{\rm\texttt {X1}} are tuples of variables, {\rm\texttt {e(X,X1)}}
is a constraint,  and {\rm\texttt {Q(X1)}} is an atom.
Then, the specialization algorithm introduces a new definition clause $D$:
{\rm $\texttt{newq(X1)\,:-\,g(X1),\,Q(X1)}$},
such that$:$ {\rm{(i)}}~{\rm\texttt{newq}} is a new predicate symbol, and
{\rm{(ii)}}~\texttt{g(X1)} is a generalization of \texttt{e(X,X1)}, that is,
 {\rm$\texttt{e(X,X1)} \sqsubseteq \texttt{g(X1)}$} (for the first
definition introduction step \texttt{g(X1)} is the {\em projection}, in the reals,
of \texttt{e(X,X1)} onto the variables \texttt{X1}). 
Then clause $E$ is folded using $D$, hence deriving the clause $F$:
{\rm$\texttt{H(X)\,:-\,e(X,X1),\,newq(X1)}$}.
Note that, even if  \texttt{g(X1)} is a generalization of \texttt{e(X,X1)},
clause $E$ is {\em equisatisfiable} to the pair of clauses $D$ and $F$. 

The clauses in \textit{Defs} are then processed similarly to
the clauses for \texttt{unsafe}, by applying unfolding,
adding new predicate definitions in \textit{Defs}, and folding, and this
unfolding-definition introduction-folding cycle is repeated until
all clauses derived by unfolding can be folded using clauses introduced in
\textit{Defs} in a previous step, so that no new predicate definitions need be 
introduced.

The termination of the specialization algorithm depends on a strategy
that controls the introduction of new definitions so that all clauses are eventually folded.
To guarantee termination we use a generalization operator
$\mathit{Gen}$ which enjoys properties similar to the 
{\em widening operator} considered in the field of abstract interpretation~\cite{Cousot_POPL77}.
In particular, given a clause $E$ as above and a set \textit{Defs} of 
predicate definitions, $\mathit{Gen}(E,\textit{Defs})$
 is a clause {\rm $\texttt{newq(X1)\,:-\,g(X1),\,Q(X1)}$},
such that~{\rm$\texttt{e(X,X1)} \sqsubseteq \texttt{g(X1)}$} and,
moreover, any sequence of applications of $\mathit{Gen}$ {\em stabilizes}, that is,
the following property holds.
For any infinite sequence $E_1,E_2,\ldots$
of clauses, let $G_1,G_2,\ldots$ be a sequence of clauses constructed as follows$:$
$(1)$~$G_1=\mathit{Gen}(E_1,\emptyset)$,
and $(2)$~for every $i\!>\!0$, $G_{i+1}=\mathit{Gen}(E_{i+1},\{G_1,\ldots,G_i\})$.
Then there exists an index $k$ such that, for every $i\!>\!k$, $G_i$  is
equal, modulo the head predicate name, to a clause in
$\{G_1,\ldots,G_k\}$.
% Jorge: the references for the specialization literature are
% originally ~\cite{Fi&13a,LeB02,PeG02} but by the time I submitted I
% didn't have LeB02 and PeG02.
Many concrete generalization operators have been defined in the CLP
specialization literature (see, for instance,~\cite{Fi&13a}), and we
will consider two of them in our experiments of Section \ref{sec:exp}.

The correctness of the specialization algorithm directly follows from
the fact that the transformation rules preserve the least model
semantics~\cite{EtG96}. Thus, given a set $VC$ of CLP clauses 
(representing verification conditions), if
$VC'$ is the output of the specialization algorithm
applied to $VC$, then {\rm $VC \models \texttt{unsafe}$} iff
{\rm $VC' \models \texttt{unsafe}$}.

\vspace{-3mm}
\subsection{Interpolating solver}
\label{subsec:method-step4}

%% A detailed description of an IHC solver is beyond the
%% scope of this paper. Instead, we will describe informally how the IHC
%% solver considered in this paper works.  We assume the reader is
%% familiar with the operational semantics of CLP
%% programs~\cite{Marriott:1998:ICL:551180} and with the concept of
%% \emph{Craig} interpolation~\cite{interpolant}. For further details, we
%% refer the reader to~\cite{GangeNSSS13}.

We now describe informally how the IHC solver considered in this
paper works. For that, the concept of a derivation plays a key role.
A \emph{derivation step} is a transition from state
\statepair{$G$}{$C$} to state \statepair{$G'$}{$C'$}, written
\statepair{$G$}{$C$} $\Rightarrow$ \statepair{$G'$}{$C'$}, where $G$,
$G'$ are goals (sequences of literals that can be either atoms or
constraints) and $C$, $C'$ are \emph{constraint stores} (the
constraints accumulated during the derivation of a goal).
A derivation step consists essentially of \emph{unifying} 
some atom in 
the current goal $G$ with the head of some clause and replacing $G$ with the
literals in the body of the matched clause producing a new goal
$G'$. Moreover, new constraints can be added to the constraint store
$C$ producing a new constraint store $C'$. At any derivation step the
constraint store $C$ can be unsatisfiable, hence producing a
\emph{failed derivation}.
A \emph{derivation tree} for a goal $G$ is a tree with states as nodes
where each path corresponds to a possible derivation of $G$.

In order to prove that a goal (e.g., the predicate \textsf{unsafe})
is unsatisfiable, the solver tries to produce a finite derivation
tree proving that the goal has no answers (i.e., all the
derivations fail). If an answer is found then it represents a
counterexample. 
%The main goal of the IHC solver is to produce a finite derivation
%tree while proving that there is no solutions (i.e., all the
%derivations fail). If a solution is found then it is a real
%counterexample. 
To facilitate this process, each node in the tree is
annotated with an interpolant producing at the end a \emph{tree
  interpolant}. To achieve this, the solver computes a
\emph{path interpolant} from each failed derivation and then combines
them.
Informally, given a sequence of formulas $F_1,\ldots,F_n$ (extracted
from the constraint store at each state in the failed derivation) the
sequence $I_0,\ldots,I_n$ is called a {path interpolant} if, for
all $i \in [1,\ldots,n]$, we have $I_{i-1} \wedge F_i \models I_{i}$
(with $I_0=true$ and $I_n=false$) and the variables of $I_i$ are common
to the variables of $F_i$ and $F_{i+1}$. The interpolant associated
with a node in the tree is the \emph{conjunction} of the children's
interpolants.

Even if the derivation tree is finite, its tree interpolant is very
valuable since it can be used for pruning other redundant failed
derivations. 
%We explain this phenomenon through an example in Section~\ref{sec:conclusions}.

A more interesting fact is how we can use interpolants to prove that
the derivation of a goal will fail infinitely.
We rely on the same principle followed by tabled CLP in order to
subsume states in presence of infinite derivations. Whenever a cycle
is detected its execution is frozen\footnote{The freeze of an
  execution can be done in several ways. The algorithm described
  in~\cite{GangeNSSS13} performs a counter instrumentation similar
  to~\cite{McMillan10} in order to make finite the execution and
  produce interpolants.}  to avoid running infinitely, and a
backtracking to an ancestor choice point occurs. 
By repeating this, the execution of a goal will always
terminate and a tree interpolant can be computed. After completion of
a subtree, the tabling mechanism will attempt at proving that the
state where the execution was frozen can be subsumed by any of its
ancestors using an interpolant as the subsumption condition. If it
fails then its execution is re-activated and the process continues. Of
course, the execution might run forever.
The subsumption test is described informally as follows. Let $G_a$ and
$G_d$ be two atoms with the same functor and arity,
where $G_a$ is the head and $G_d$ is the tail of a cycle. The symbols
$a$ and $d$ refer to ancestor and descendant, respectively. Let
$\qquote{p_1},\ldots ,\qquote{p_n}$ be all the constraints originated
from all paths $p_1,\ldots ,p_n$ between $G_a$ and $G_d$, and $I_a$ be
the interpolant computed for $G_a$. Then, we do not need to
re-activate the execution of $G_d$ if $\bigwedge_{1 \leq i \leq n} I_a
\wedge \qquote{p_i} \models I_a'$ (where $I_a'$ is the interpolant
$I_a$ after proper renaming). This process is analogue to tabled CLP's
\emph{completion check}.

\begin{figure}[h]
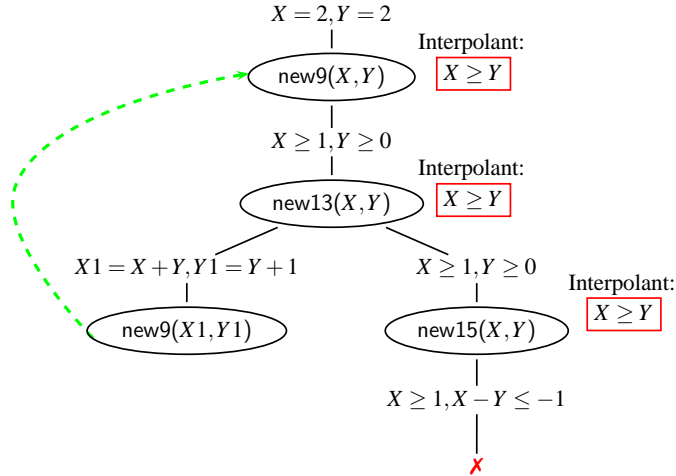

\begin{center}
 \scalebox{0.8}{
\psset{dash=4pt 4pt}

% These coordinates are not preserved if the figure is scaled. I need
% to fix this. Meanwhile if scalebox changes the ratio we need to
% changes these coordinates.

\rput[l](8.2,-4.5){Interpolant:}
\rput[l](8.5,-5.0){\psframebox[linecolor=red]{$X \geq Y$}}  %% new15
\rput[l](5.7,-2.6){Interpolant:}
\rput[l](6.0,-3.1){\psframebox[linecolor=red]{$X \geq Y$}}  %% new13
\rput[l](5.7,-0.5){Interpolant:}
\rput[l](6,-1.0){\psframebox[linecolor=red]{$X \geq Y$}}  %% new9
\pscurve[linecolor=green, linestyle=dashed, linewidth=1.5pt]{->}(0.3,-5.4)(-1.0,-2.9)(2.9,-1.0)
%\pscurve[linecolor=green, linestyle=dashed, linewidth=1.5pt]{->}(-3.3,-5.4)(-5.0,-3.5)(-1.2,-1.6)

%% For no scale:
%% rput[l](4.2,-6.4){Interpolant:}
%% \rput[l](4.5,-7.0){\psframebox[linecolor=red]{$X \geq Y$}}  %% new15
%% \rput[l](1.8,-3.9){Interpolant:}
%% \rput[l](2.0,-4.5){\psframebox[linecolor=red]{$X \geq Y$}}  %% new13
%% \rput[l](1.7,-1.4){Interpolant:}
%% \rput[l](1.9,-2.0){\psframebox[linecolor=red]{$X \geq Y$}}  %% new9
%% \pscurve[linecolor=green, linestyle=dashed, linewidth=1.5pt]{->}(-3.3,-5.4)(-5.0,-3.5)(-1.2,-1.6)

\pstree[nodesep=2pt,levelsep=30pt,treesep=10pt]{\TR{$X = 2, Y= 2$}}{
  \pstree{\Toval{\clppred{new9}{X,Y}}}{
    \pstree{\TR{$X \geq 1, Y \geq 0$}}{
      \pstree[treesep=-11pt]{\Toval{\clppred{new13}{X,Y}}}{
        % left child of p2
        \pstree{\TR{$X1=X+Y, Y1=Y+1$}}{
          \pstree[levelsep=16pt,treesep=-11pt]{\Toval{\clppred{new9}{X1,Y1}}}{
          }
        }
        \tspace{2cm}
        % right child of p2
        \pstree{\TR{$X \geq 1, Y \geq 0$}}{
          \pstree[levelsep=16pt]{\Toval{\clppred{new15}{X,Y}}}{
            \skiplevel{
              \pstree{\TR{$ X \geq 1, X-Y \leq -1$}}{
                \skiplevel{\TR{\crossmark}}
              }
            }
          }
        }
      }
    }
  }
}
}
\end{center}
  \caption{The derivation tree for goal $X=2,Y=2, \clppred{new9}{X,Y}$
    where \textsf{new9} is a predicate defined 
    in the CLP program of Figure~\ref{example}(c)}
\label{fig:example-fig}
\end{figure}

Consider again the transformed program in Figure~\ref{example}(c). Let
us focus on the execution of \textsf{new9}, which is a recursive
predicate. Recall that \textsf{new9} is reached after unwinding twice the
loop. Therefore, before the execution of \textsf{new9} the
\emph{constraint store} is $X=2, Y=2$, after constraint
simplification.
Its depth-first, left-to-right derivation tree is shown in
Figure~\ref{fig:example-fig}. Each oval node represents the call to a
body atom and an edge denotes a derivation step. A failed derivation
is marked with a (red) ``cross'' symbol. Note that there is no
successful derivation, otherwise the program would be unsafe.

The leftmost derivation is frozen when the atom \clppred{new9}{X1,Y1}
is encountered. Then, the execution backtracks to \clppred{new13}{X,Y}
and activates the rightmost derivation, which fails. 
We compute the interpolant for this derivation
and we annotate the atom
\clppred{new9}{X,Y} with the constraint $X \geq Y$. At this point, we
visit again \clppred{new9}{X1,Y1} in order to perform the subsumption
test: $ X \geq Y \wedge X \geq 1 \wedge Y \geq 0 \wedge X1 = X+Y
\wedge Y1 = Y+1 \models X1 \geq Y1$. This entailment holds, and
therefore the execution can safely stop proving that the goal
\clppred{new9}{X,Y} is unsatisfiable. Note that if the transformation had
not added the constraints $X \geq 1, Y \geq 0$ the subsumption test
would have failed and the execution would have run forever.

Finally, the criteria used for stopping the IHC solver is currently
based on a timeout. Of course, due to undecidability reasons, there is
no method that can decide whether the IHC solver will eventually stop
finding a safe inductive invariant. However, there might be cases
where by inspecting the interpolants we could guess that it is not
likely for the IHC solver to stop in a reasonable amount of
time. In these cases, it
is desirable to switch to the next transformation phase
instead of waiting until the timeout expires. In
the future, we would like to investigate this problem.

%\vspace{-3mm}
\subsection{CLP reversal}
\label{subsec:method-step5}

In the case where the IHC
solver is not able to check (within a given amount of time)
whether \texttt{unsafe} holds or not, 
the verification method returns to Step 3 
and propagates the constraints by first inverting the roles of
the initial and error configurations.
Thus, at each iteration of the method, verification
switches from forward propagation (of the constraints of the initial configuration)
to backward propagation (of the constraints of the error configuration), or vice versa,
having also strengthened the constraints of the initial and error configurations
due to previous specializations.
This switch is achieved by a CLP transformation called {\em Reversal}~\cite{De&14c}.
 
CLP Reversal transforms the set $\mathit{VC1}=\mathrm{\{1,2,3\}}$ of CLP clauses
into the set $\mathit{VC2}=\mathrm{\{4,5,6\}}$.

\smallskip

{\small

\begin{tabular}{ l   l }

1.~~$\mathtt{unsafe}$ {\tt{:-}} 
$\mathtt{a(U), ~r1(U).}$
&\hspace*{2cm}
4.~~$\mathtt{unsafe}$ {\tt{:-}} $\mathtt{b(U), 
~r2(U).}$
\\

2.~~$\mathtt{r1(U)}$ {\tt{:-}} 
$\mathtt{c(U,V), ~r1(V).}$
&\hspace*{2cm}
5.~~$\mathtt{r2(V)}$ {\tt{:-}} 
$\mathtt{c(U,V), ~r2(U).}$

\\

3.~~$\mathtt{r1(U)}$ {\tt{:-}} 
$\mathtt{b(U).}$
&\hspace*{2cm}
6.~~$\mathtt{r2(U)}$ {\tt{:-}} 
$\mathtt{a(U).}$
\\

\end{tabular}

}

\smallskip
\noindent
The Reversal transformation can be generalized to
any number of clauses and predicates, and preserves safety in the sense that
$\mathit{VC1} \models \mathtt{unsafe}$ iff
$\mathit{VC2} \models \mathtt{unsafe}$.

%\end{document}

%\documentclass[14-HCVS-FLoC.tex]{subfiles}
%\begin{document}

\section{Experimental Evaluation}  
\label{sec:exp}

%\emanuele{DRAFT1: logical structure, names 
%to be updated according to those defined in Section 3} 

The verification method presented in Section~\ref{sec:method} has been
implemented by combining \mbox{VeriMAP}~\cite{DeAngelisFPP14} and
FTCLP~\cite{GangeNSSS13}.  The verification process is controlled by
VeriMAP, which is responsible for the orchestration of the following
components: (i)~a \textit{translator}, based on the C Intermediate
Language (CIL)~\cite{NeculaMRW02}, which translates a given
verification problem (i.e., the C program together with the initial
and error configurations) into a set of CLP program, (ii)~a
\textit{specializer} for CLP programs, based on the MAP transformation
system~\cite{MAP}, which generates the verification conditions (VCs)
and applies the iterated specialization strategy, and (iii)~an {\em
  IHC solver} ($\textit{IHCS}$), implemented by the FTCLP tool.

%% In order to evaluate our software model checking method we have
%% performed an experimental evaluation on several benchmark programs
%% taken from the literature.  The benchmark set used in our experiments
%% consists of 216 verification problems (179~of which are safe, and the
%% remaining~37 are unsafe).  

%% Most problems have been taken from the benchmark sets of other tools
%% used in software model checking, like DAGGER~\cite{DAGGER08} (21
%% problems), TRACER~\cite{JaffarMNS12} (66 problems) and
%% InvGen~\cite{GuptaR09} (68 problems), and from the TACAS~2013 Software
%% Verification Competition~\cite{Beyer13} (52 problems).  The size of
%% the input programs ranges from a dozen to about five hundred lines of
%% code.  The source code of all the verification problems we have
%% considered is available at
%% {\tt{http://map.uniroma2.it/VeriMAP/hcvs/}}.

We have performed an experimental evaluation on a set of benchmarks
consisting of 216 verification problems (179~of which are safe, and
the remaining~37 are unsafe).
Most problems have been taken from the repositories of other tools
such as DAGGER~\cite{DAGGER08} (21 problems),
TRACER~\cite{JaffarMNS12} (66 problems), InvGen~\cite{GuptaR09} (68
problems), and also from the TACAS~2013 Software Verification
Competition~\cite{Beyer13} (52 problems).  The size of the input
programs ranges from a dozen to about five hundred lines of code.  The
source code of all the verification problems is available at
{\tt{http://map.uniroma2.it/VeriMAP/hcvs/}}.

The program verifier has been configured to execute the
following process:

\smallskip

$\textit{Specialize}_{\textit{Remove}} ;~
\textit{Specialize}_{\textit{Prop}} ;~
\textit{IHCS} ;~
\big(\textit{Reverse} ; ~\it{Specialize}_{\textit{Prop}} ;~\textit{IHCS}\big)^{*}$

\smallskip
\noindent
After having translated the verification problem $P$ into CLP, the
verifier: (i) generates the verification conditions for $P$ by
applying the $\textit{Specialize}_\textit{Remove}$ procedure
(Section~\ref{subsec:method-step2}), (ii) propagates the constraints
that represent the initial configurations by executing the
$\textit{Specialize}_\textit{Prop}$ procedure
(Section~\ref{subsec:method-step3}), and (iii) runs the \textit{IHC}
solver (Section~\ref{subsec:method-step4}).
If the solvability of the CLP clauses can be decided the verifier
stops. Otherwise, the verifier calls the \textit{Reverse} procedure
that interchanges the roles of the initial and error configurations
(Section \ref{subsec:method-step5}), and calls 
$\textit{Specialize}_\textit{Prop}$ again. The $\big(\textit{Reverse} ;
~\it{Specialize}_{\textit{Prop}} ;~\textit{IHCS}\big)$ sequence might
repeat forever unless the specializer is able to generate a set of CLP
clauses that \textit{IHCS} can either prove to be solvable or prove to
be unsolvable.

All experiments have been performed on an Intel Core Duo E7300 2.66Ghz
processor with 4GB of memory under the GNU Linux operating system
Ubuntu 12.10 (64 bit, kernel version 3.2.0-57).  A timeout limit of
two minutes has been set for each verification problem.

Table~\ref{tab:experiments} summarizes the verification results
obtained by the VeriMAP and the FTCLP tools executed separately
(first, second and fourth columns) and the combination of both tools
(third and fifth columns). When VeriMAP is executed without the help of
FTCLP, the analysis described in~\cite{De&14c} is used in
place of the IHC solver.
In the columns labeled by VeriMAP$_\textit{M}$ and
VeriMAP$_\textit{PH}$ we have reported the results obtained by using
the VeriMAP system with the generalization operator
$\textit{Gen}_\textit{M}$ ({\em monovariant
  generalization}\footnote{All constrained atoms with the same
  predicate are generalized to the same new predicate.}  using
widening only) and $\textit{Gen}_\textit{PH}$ ({\em polyvariant
  generalization}\footnote{Constrained atoms with the same predicate
  can be generalized to different new predicates.} using widening and
convex hull), respectively.
Row~1 reports the total number of definite answers 
(correctly asserting either
program {\em safety} or {\em unsafety}).  Row~2 reports the number of
tool crashes.  Row~3 reports the number of verification problems that
could not be solved within the timeout limit of two minutes.  Row~4
reports the total CPU time, in seconds, taken to run the whole set of
verification tasks: it includes the time taken to produce answers and
the time spent on tasks that timed out. Finally, row~5 reports the
average time needed to produce a definite answer, which is obtained by
dividing the total time by the number of answers.

%% The {\textit{Specialize}}$_{\textit{Prop}}$ procedure has been
%% executed by using the four generalization operators presented in
%% Section~\ref{sec:Verification}: (i)~$\textit{Gen}_\textit{M}$, that is
%% a monovariant generalization with widening only,
%% %(ii)~$\textit{Gen}_\textit{MH}$, that is a monovariant generalization
%% %with widening and convex hull,
%% %(iii)~$\textit{Gen}_\textit{P}$, that is a polyvariant generalization with widening only, and
%% (ii)~$\textit{Gen}_\textit{PH}$, that is polyvariant generalization
%% with widening and convex hull.

\begin{table*}[t]
\begin{small}
\centering
\begin{tabular}{|@{\hspace{4pt}}l@{\hspace{4pt}}||@{\hspace{4pt}}r@{\hspace{4pt}}|@{\hspace{4pt}}r@{\hspace{4pt}}|
@{\hspace{4pt}}r@{\hspace{4pt}}|@{\hspace{4pt}}r@{\hspace{4pt}}|@{\hspace{4pt}}r@{\hspace{4pt}}|}\hline
& FTCLP &	 VeriMAP$_{M}$	& VeriMAP$_{M}$\,+\,FTCLP & VeriMAP$_{PH}$  &	VeriMAP$_{PH}$\,+\,FTCLP\\
\hline	
answers & 116	& 128& 	160 	& 178 & 	182\\
\hline	
crashes & 5 & 0	& 2 & 	0 & 0\\
\hline	
timeouts & 95 & 88	& 54 & 38 & 	34\\
\hline	
total time	& 
12470.26	& 11285.77 &	 9714.41	& 5678.09 &	6537.17\\
\hline	
average time	& 107.50 & 	88.17 &	60.72 &	31.90	& 35.92\\
\hline	
\end{tabular}
\caption{Verification results using VeriMAP, FTCLP, and the combination of 
VeriMAP and FTCLP. The timeout limit is two minutes. Times are in seconds.}
\label{tab:experiments}
\end{small}
\end{table*}

%+++ EMA: benefits of widening may be misleading, we use generalization
%operator based on widening (and, in the experiments we also use convex hull).
%Those operators play different roles in the verification process base on 
%program specialization. 
%I would prefer to make a clear distinction between
%generalization operators based on widening and widening operators used
%in the abstract interpretation.
%Moreover, I would like to emphasize the role of constraint propagation
%performed by program specialization which is, in my opinion, what makes
%the analysis phase more effective.
%+++

The results in Table~\ref{tab:experiments} show that the combination
of VeriMAP and FTCLP, by exploiting the synergy of widening and
interpolation, improves the performance of both tools whenever
executed separately.  In particular, we have that the best performance
is achieved by the combination VeriMAP$_{PH}$ + FTCLP where the
process is able to provide an answer for 182 programs out of
216 ($84.26\%$).

Table~\ref{tab:experiments_ite} summarizes the results obtained at the
end of each of the first five iterations of the verification process
when VeriMAP is executed alone and combined with FTCLP.  We observe
that when VeriMAP is used in combination with FTCLP the number of
iterations required to solve the verification problems is considerably
reduced.

\begin{table*}[t]
\begin{small}
\centering
\begin{tabular}{|c||r|r|r|r|}\hline

Iteration  &	  VeriMAP$_{M}$ & VeriMAP$_{M}$ + FTCLP &	 VeriMAP$_{PH}$ & VeriMAP$_{PH}$ + FTCLP\\
  \hline
1 &	 74	 &	119 &	104	 &	136\\
\hline
2 & 45	 &	38 & 54 &	 34\\
\hline
3 &	 7 & 2 &	10 & 5\\
\hline
4 &	 2 & 1 &	8  & 3\\
\hline
5 &	 0	& 0 &	2  & 4 \\
\hline

\end{tabular}
\caption{Number of definite answers computed by VeriMAP and by the
  combination of VeriMAP and FTCLP within the first five iterations.}
\label{tab:experiments_ite}
\end{small}
\end{table*}

\section{Related Work}
\label{sec:related}

As Horn logic is becoming more popular for reasoning about properties of
programs, the number of verifiers
based on this logic has increased during recent years (e.g,~\cite{DeAngelisFPP14,GrebenshchikovLPR12,HoderBM11,JaffarMNS12,Duality,disjunctive-intps,ViG14}). 
Although they
can differ significantly from each other, one possible classification
is based on their use of interpolation
(e.g.,~\cite{GrebenshchikovLPR12,JaffarMNS12,Duality,disjunctive-intps}),
Property Directed Reachability (PDR)~\cite{Bradley11}
(e.g.,~\cite{HoderBM11}), or a combination of both (e.g.,~\cite{ViG14}).
Unlike the above mentioned verifiers, VeriMAP~\cite{DeAngelisFPP14} does
not use interpolation and, as explained in previous sections,
implements a transformation method based on widening techniques
similar to the ones used in the field of abstract
interpretation~\cite{Cousot_POPL77}.

It should be noted that, with the exception of VeriMAP, abstract
interpretation techniques are surprisingly less common than PDR and
interpolation in Horn Clause verifiers. HSF~\cite{GrebenshchikovLPR12}
combines predicate abstraction with interpolation but no other
abstract interpretations. TRACER~\cite{JaffarMNS12} only uses abstract
interpretation as a pre-processing step in order to inject invariants
during the execution of the Horn Clauses.  Therefore, to the best of
our knowledge there is no Horn Clause verifier that combines abstract
interpretation (apart from predicate abstraction) with interpolation
in a nontrivial manner.

%% For the rest of this section we focus on combinations of abstract
%% interpretation with interpolation outside the scope of Horn Clause
%% verifiers.

%% \noindent
%% \vspace{2mm} \textbf{Abstract Interpretation and
%%   Interpolation}. Counterexample driven refinement
%% (CEGAR)~\cite{ClarkeGJLV00} has been a very effective method to solve
%% reachability problems in infinite-state systems. Among many
%% improvements, two major techniques have had a great influence in its
%% success: \emph{predicate abstraction}~\cite{graf97predicate} for
%% producing precise but yet practical abstractions, and
%% \emph{interpolation}~\cite{Henzinger2004} for refining succinctly
%% coarse abstractions.

Several works (e.g.,~\cite{AGCSAS12,DAGGER08,GulavaniR06,WangYGI07})
have focused on how to refine abstract interpretations different from
predicate abstraction outside the scope of Horn Clause verifiers.  %% The
%% two major sources of imprecision are due to the use of widening at
%% loopheads and join operators at locations where conditional branches
%% merge. 
%
\cite{GulavaniR06,WangYGI07} focus on how to recover from the losses
of widening by using specific knowledge of
polyhedra. \textsc{Dagger}~\cite{DAGGER08} tackles in a more general way the
imprecision due to widening by proposing the ``interpolated widen''
operator which refines the abstract state after widening using
interpolation. \cite{AGCSAS12} proposes another algorithm called \textsc{Vinta}
which can also refine precision losses from widening, but it relies
heavily on the use of an abstract domain that can represent
efficiently disjunction of abstract states. However, efficient
disjunctive abstract domains are rare and it is well known that the
design of precise widening operators is far from
easy. 
\textsc{Ufo}~\cite{AlbarghouthiGC12,AlbarghouthiLGC12}
is a framework for combining CEGAR methods based on over-and-under
approximations which is parameterized by a \emph{post} operator. The
post operator is used during the unwinding of the control flow graph. 
If an error is found then Craig interpolation is used to refine the
abstraction. Although in principle the post operator could perform an
arbitrary abstraction the refinement described
in~\cite{AlbarghouthiGC12,AlbarghouthiLGC12} assume heavily that
predicate abstraction is used. If other abstractions were used it is
not clear at all how to refine them. Moreover, 
unlike \textsc{Dagger} and \textsc{Vinta}, during the unwinding of
the control flow graph no abstract joins are performed, and thus we may
consider \textsc{Ufo} as another CEGAR method based on interpolation.

%% Longer version:
%% The DAGGER algorithm~\cite{DAGGER08} tackles in a more general way the
%% imprecision due to widening and join operators by proposing two
%% somewhat orthogonal ideas: (a) an ``interpolated widen'' operator
%% which refines the abstract state after widening using interpolation,
%% and (b) a new algorithm that splits explicitly joined states using
%% also interpolation. \cite{AGCSAS12} proposes another algorithm called
%% VINTA which in addition to (a) and (b) can also refine losses from the
%% limitations of the abstract semantics. VINTA relies heavily on the use
%% of an abstract domain that can represent efficiently disjunction of
%% abstract states. As a result, VINTA does not suffer from an explicit
%% blow up as DAGGER when it splits merged states. However, efficient
%% disjunctive abstract domains are rare and it is well known that the
%% design of precise widening operators is far from easy.

The approach followed by \textsc{Dagger} is probably the most closely related to
ours. If \textsc{Dagger} finds a spurious counterexample due to widening losses,
it must be the case that $(A \lub B) \meet E = \bot$ but $(A \widen B)
\meet E \neq \bot$, where $A$ is the abstract state before starting
the execution of a loop, $B$ is the abstract state after executing the
backedge of the loop, and $E$ is an abstract state that leads to an
error. The idea behind ``interpolated widen'' is to replace $\widen$
with $\widenupto{I}$, where $\widenupto{I}$ is an instance of the
widening up-to~\cite{HalbwachsPR97}. The key property of the
$\widenupto{I}$ operator is that it preserves the desirable properties
of widening while excludes from the abstract state $A \widenupto{I} B$
the spurious counterexample (and possibly others) denoted by the
interpolant $I$ (i.e., $(A \widenupto{I} B) \meet E = \bot$).
Our transformation phase performs widening during the generalization
step, while the IHC solver generates interpolants in order to discover
more program invariants. We believe this combination can be seen as a
version of the $\widenupto{I}$ operator. The main difference is that
our method can obtain an effect similar to combining widening with
interpolation without the enormous effort of implementing a new
verifier from scratch.

%\end{document}

%\documentclass[14-HCVS-FLoC.tex]{subfiles}
%\begin{document}

\section{Conclusions and Future Work}
\label{sec:conclusions}

In this paper we have presented some preliminary results obtained by
integrating an Interpolating Horn Clause solver (FTCLP) with an
Iterated Specialization tool (VeriMAP). The experimental evaluation
confirms that such an integration is effective in practice, as
discussed in Section~\ref{sec:exp}.

The fact that both tools use CLP as a representation formalism
for the verification conditions, together with the modular design of
VeriMAP, allowed us a very clean and painless integration with
FTCLP. As a result, we can achieve the effect of combining abstract
interpretation with interpolation without having to design and
implement a custom verifier. We believe this modular combination is
valuable by itself, since based on the experience, one could implement
a custom verifier or simply apply the method described here if the
performance is adequate.

In this preliminary work, we have used the IHC solver mainly as a
black-box and although the gains are promising they are somewhat
limited. As future work, we would like to combine these tools in more
\emph{synergistic} ways. We believe that the integration can be
improved in several ways.

First, when FTCLP is not able to produce a solution within the
considered timeout limit, it would be useful to leverage the partial
information it discovers and integrate it in the transformed program,
with the aim of improving the subsequent unfold/fold transformation
steps.
For example, during its execution FTCLP might discover that some
subtrees rooted in some goal cannot lead to an answer.  Thus,
the corresponding predicate can be considered useless and its clauses
can be removed from the specialized
program before the next iteration starts.

Another observation is that FTCLP generates for each predicate $p$ an
interpolant that represents an over-approximation of the original
constraint store that preserves the unsolvability of $p$.  It would
be interesting to study how these interpolants can be used to refine
the generalization step performed during the unfold/fold
transformation, with the objective of preserving the branching
structure of the symbolic evaluation tree, as indicated
in~\cite{DeAngelisFPP12b}, and preventing the introduction of
spurious paths.

Finally, another possible direction for future work regards the use of
interpolation \emph{during} the transformation process in order to
make more efficient the unfold/fold process.  While this appears to be
a very promising direction it raises some issues related to the
termination of the transformation process itself, which deserve
further study.

\section*{Acknowledgments}

We would like to thank the anonymous referees for their helpful
and constructive comments.
This work has been partially supported by the Italian
National Group of Computing Science (GNCS-INDAM).

\bibliographystyle{eptcs}

\bibliography{refs}

%\subfile{99biblio.tex}

\end{document}